%% file: main.tex
\documentclass[conference]{IEEEtran}
\IEEEoverridecommandlockouts
\def\BibTeX{{\rm B\kern-.05em{\sc i\kern-.025em b}\kern-.08em
    T\kern-.1667em\lower.7ex\hbox{E}\kern-.125emX}}

\pdfoutput=1

\usepackage{cite}
\usepackage{amsmath,amssymb,amsfonts}
\usepackage{graphicx}
\usepackage{textcomp}
\usepackage{xcolor}
\usepackage{url}
\usepackage[english]{babel}   
\usepackage[utf8]{inputenc}
\usepackage{amsthm}

\usepackage[english,        % Language
            algoruled,      % Places horizontal rules to delimit the algorithm box
            noline,         % Don't place vertical lines to mark code blocks
           ]{algorithm2e}   % For writing algorithms
\usepackage[framemethod=TikZ]{mdframed} % For making colored boxes around text
\usepackage[
    colorlinks=true,linkcolor=blue,citecolor=blue,
    filecolor=magenta,urlcolor=blue
]{hyperref}

\newtheoremstyle{elfthmstyle}
  {3pt} % Space above
  {1.5pt} % Space below
  {} % Body font
  {} % Indent amount
  {\bfseries} % Theorem head font
  {.} % Punctuation after theorem head
  {.5em} % Space after theorem head
  {} % Theorem head spec (can be left empty, meaning `normal')

\theoremstyle{elfthmstyle}
\newtheorem{elfproposition}{Proposition}

\allowdisplaybreaks

\mdfsetup{
    middlelinecolor=blue!30,
    middlelinewidth=2pt,
    backgroundcolor=cyan!5,
    roundcorner=20pt,
    nobreak=true
}

\begin{document}

\title{High Performance Algorithms for Counting Collisions and Pairwise Interactions
\thanks{This paper has been accepted in the International Conference on Computational Science (ICCS2019) and published in Springer's Lecture Notes in Computer Science series. The final authenticated version is available online at: \url{https://doi.org/10.1007/978-3-030-22734-0_14}.}}

\author{
    \IEEEauthorblockN{Matheus Henrique Junqueira Saldanha}
    \IEEEauthorblockA{
        % \textit{LaSDPC -- SSC} \\
        %\textit{Institute of Mathematics and} \\ 
        %\textit{Computer Science} \\
        %\textit{University of São Paulo} \\
        %São Carlos, SP, Brazil\\
        mhjsaldanha@gmail.com
    }
    \and
    \IEEEauthorblockN{Paulo Sérgio Lopes de Souza}
    \IEEEauthorblockA{
        % \textit{LaSDPC -- SSC} \\
        %\textit{Institute of Mathematics and} \\ 
        %\textit{Computer Science} \\
        %\textit{University of São Paulo} \\
        %São Carlos, SP, Brazil\\
        pssouza@icmc.usp.br
    }
    \and
    \IEEEauthorblockN{\hspace{0cm}}
    \and
    \IEEEauthorblockA{
        \textit{\hspace{5.5cm} Institute of Mathematics and Computer Sciences} \\
        \textit{\hspace{5.5cm} University of São Paulo -- São Carlos, SP, Brazil}
    }
}

\maketitle

\begin{abstract}
% collision counting
% pairwise interaction
% GPU
% parallel
% efficient
% high performance
The problem of counting collisions or interactions is common in areas as computer graphics and scientific simulations. Since it is a major bottleneck in applications of these areas, a lot of research has been carried out on such subject, mainly focused on techniques that allow calculations to be performed within pruned sets of objects.
% determine smaller sets of objects among which to calculate interactions with the usual brute-force algorithm.
%Paulo: sugiro tirar essa frase abaixo. Acho que ela deprecia e levanta questionamentos.Se for necessária, precisa ser reescrita para algo como "The main focus in our paper is improve usual algorithms found in the literature."
%This paper focuses on the brute-force algorithm itself.
This paper focuses on how interaction calculation (such as collisions) within these sets can be done more efficiently than existing approaches. Two algorithms are proposed: a sequential algorithm that has linear complexity at the cost of high memory usage; and a parallel algorithm, mathematically proved to be correct, that manages to use GPU resources more efficiently than existing approaches. The proposed and existing algorithms were implemented, and experiments show a speedup of 21.7 for the sequential algorithm (on small problem size), and 1.12 for the parallel proposal (large problem size).
% Computer graphics, virtual reality, robotics, simulations
% People are connected with each other through social network, internet, games...
% Things are connected
% By improving interaction calculation, this work contributes to technologies that connect people (such as games and virtual reality) and robots (collision-free path detection).
% \rev{Interconnected world towards which we are heading, which is pervaded by objects that sense the space around them. Or objects with spatial senses/perception. Or objects that can perceive the surrounding environment.}
By improving interaction calculation, this work contributes to research areas that promote interconnection in the modern world, such as computer graphics and robotics.
\end{abstract}


\input{text/1-introduction.tex}
\input{text/2-relatedwork.tex}
\input{text/3-linearapproach.tex}
\input{text/4-parallelalgorithm.tex}

\input{text/5-experiments.tex}

\input{text/6-conclusions.tex}
\input{text/9-acknowledgements.tex}

\bibliographystyle{ieeetr}
\bibliography{references}

% \appendix
% \input{text/A-proof-even-N.tex}

\end{document}

%% file: text/1-introduction.tex
\section{Introduction}
\label{section:1-introduction}

As the performance growth of a single processor core decreases, attention has been shifting towards other means to decrease execution time of programs. Given the lower price of primary and secondary memory, compared to processors, exchanging memory usage for reduction on algorithm complexity is an interesting option. Other commonly explored alternative is to exploit the increasing parallelism available in hardware, which can be accomplished by parallelizing existing sequential algorithms. This does not always work well, and a good parallel algorithm might only be brought forth by complete re-analysis of the problem and design of new parallel algorithms. In this article, we provide solutions that explore these two options to accelerate the problem of calculating interactions (e.g. forces, contacts) in N-body environments.
The problem of interaction counting is widely present in areas that promote interconnection in the modern world: computer graphics and virtual reality, that connect people through games and link computer science to medicine, for example, where virtual environments can be used for practicing surgery; robotics, that interconnect objects and robots; and scientific simulations, that connect computing with areas that need to simulate complex natural phenomena such as protein folding and planet motion. Despite being largely used, interaction counting is a major bottleneck\cite{lin1998collisionsurvey} in many applications. It may come in two flavors: 1) interactions might only matter within subsets of objects (e.g. collisions, where only neighbor objects are relevant), and it is common to use strategies such as spatial partitioning\cite{elseberg2012comparison-nnalgs} or bounding volume hierarchies\cite{stich2009spatial-bvh} to find these subsets before performing interaction counting; or 2) all interactions matter (e.g. gravitational forces), where it is common to use techniques that allow a set of objects to be regarded as a single object, such as the Fast Multipole Method\cite{greengard1987fastmultipole}. Besides that, it is also often the case where the objects in question are in motion and collision detection must be done every small time steps, such as in a robot calculating collision-free paths. For such situations, beside the aforementioned strategies, time and space coherence is also used, that is, the fact that objects won't move too much in a short time span is exploited\cite{gregory2005framework}. In any of these cases, however, there will still be a phase where smaller sets of objects undergo interaction calculation, and the usual way is to iterate over each object and compare it with the others, giving a $O(N^2)$ complexity.

Even though a lot of research has been carried out on alternative strategies to amortize the cost of this pairwise-comparisons $O(N^2)$ approach, not much has been done regarding this brute-force algorithm itself. In this paper is proposed a new parallel algorithm for calculating interactions that is designed to make better use of architectural resources on GPUs. The algorithm is mathematically proved to be correct, and results show a speedup of 1.12 over parallelization of the straightforward approach. We also propose a sequential algorithm for counting collisions among punctual objects, which has a $O(N)$ complexity at the cost of high virtual memory usage, and experiments show a speedup of 21.7 for limited size objects. By accelerating the pairwise-comparisons algorithm, we hope to facilitate smoother animations and virtual reality environments, simulations that take less time (or provide better results in the same amount of time), and robots that can better avoid collisions with its surroundings.
%; and these are important entities in a world where people and objects are interconnected, especially robots, animations (for games) and virtual reality.

% The document is organized as follows. There has been a large amount of research in the area of collision counting, which is discussed in Section \ref{section:2-relatedwork}. In Sections \ref{section:3-linearapproach} and \ref{section:4-parallelalorighms} two algorithms for collision counting are proposed; one is sequential and trades execution time for memory usage, while the other is parallel and has interesting properties that make it efficient on GPUs. Experiments performed for evaluating the proposed algorithms are shown and discussed in \ref{section:5-experiments}. Finally, Section \ref{section:6-conclusion} concludes the paper.
The document is organized as follows. An overview of research in collision counting is discussed in Section \ref{section:2-relatedwork}. In Sections \ref{section:3-linearapproach} and \ref{section:4-parallelalorighms}, two algorithms for collision counting are proposed. Experiments with the proposed algorithms are discussed in Section \ref{section:5-experiments}. Finally, Section \ref{section:6-conclusion} concludes the paper.

%% file: text/2-relatedwork.tex
\section{Related Work}
\label{section:2-relatedwork}

Calculating interactions is a common problem that has been the subject of a vast amount of interesting research. Covering all of its facets is not our objective here, so in the following we provide a short overview of the subject.

I-COLLIDE \cite{cohen1995collide} is a well-known suite of algorithms that perform fast collision detection among a large number of rigid or deforming objects, such as an environment with tens of thousands of triangles (that might compose an object). They focus on pruning potentially colliding sets (PCS) multiple times before performing the exact collision detection. CULLIDE \cite{govindaraju2003cullide} follows the same line, and is parallelized for GPUs. Specializations of these algorithms for haptic systems, where collision detection must be performed thousands of times per second, were done in \cite{gregory2005framework, gregory2000six}.

% H-COLLIDE \cite{gregory2005framework} is similar to I-COLLIDE, but with a focus on haptic systems, where collision detection must be performed thousands of times per second. Besides exploiting temporal coherence inherent to haptic systems, it uses both spatial decomposition and bouding volume hierarchies to achieve good pruning of what they call the potentially colliding set.

% Counting collisions among polygonal models is studied in \cite{gregory2000six} for making haptic technology, which requires collision counting to be performed thousands of times per second, viable. They exploit temporal and spatial coherence to reject impossible collisions before performing object-object collision detection.

Spatial partitioning and bounding volumes hierarchies (BVHs) are broadly used for defining reduced sets of potentially colliding objects. A comparison among such techniques
% for finding nearest neighbors
is found in \cite{elseberg2012comparison-nnalgs}, and k-d trees and octrees are notable in such problem. Reference \cite{stich2009spatial-bvh} gives an efficient approach to construct BVH trees borrowing techniques from spatial partitioning, and \cite{wald2007fast} focuses on fast re-building of these trees on situations where there are moving objects.

% In \cite{stich2009spatial-bvh} is offered an approach to construct BVH trees more efficiently using techniques from spatial partitioning. Their approach manages to reduce overlapping volumes, hence improving performance.

% In \cite{wald2007fast} fast re-building of BVHs is studied, which can aid exploiting temporal coherence on collision detection for moving objects.

Collision detection often involves calculating trajectories of objects, which is not always simple. In \cite{tang2014fast} the authors present fast and accurate methods for evaluating collision between triangulated models in such circumstances; their method involve a series of \textit{coplanarity} and \textit{inside} tests among elements (e.g. edges, vertices). In \cite{selle2008mass}, where hair simulation is explored, there is a large number of hair-body collisions and hair-hair interactions (collisions and other forces such as friction and static attraction). In either case, the GPU approach we propose in this paper could be used to accelerate calculation of interactions, such as the coplanarity tests. Similar studies are found in \cite{provot1997collision, bridson2002robust, redon2002fast, tang2010efficient, brochu2012efficient}.

More related to engineering, in \cite{zheng2012gpu} is proposed a parallel algorithm for contact detection using spatial partitioning strategies, which is then experimented on simulation of concrete. Similarly, a framework that uses GPU for evaluating forces among particles of sand is given in \cite{longmore2013towards}. Finally, in \cite{nguyen2007gpu} the authors show an efficient parallelization, in the CUDA programming model, for the problem of evaluating gravitational forces among all bodies in an N-body system. They parallelized the straightforward sequential algorithm, where each body is compared with every other body. For summing these forces, in this paper we
% analyze the problem of calculating the sum of such forces among bodies, and
propose a different sequential algorithm whose parallelization manages to make better use of GPU resources. To the best of our knowledge, this is the first work that brings an alternative to the pairwise-comparisons approach (for summing interactions) which provides benefits when parallelized.

%% file: text/3-linearapproach.tex
\section{A Sequential \textit{O(N)} Approach}
\label{section:3-linearapproach}

% Overwrite the \S command with what we want
\let\oldS\S
\renewcommand{\S}{\mathcal{S}}

Consider the problem of counting the total number of collisions or contacts among beads (i.e. punctual objects) in a space $\S^3$ such that
$$
\begin{cases}
    \S^3 = \S \times \S \times \S
    \\
    \S = \{-a,\ -a+1,\ \dots,\ a-1,\ a\} \subset \mathbb{Z}
\end{cases}
$$
that is, $\S^3$ is a three-dimensional, discrete and finite space whose axes are symmetric around the $0$. Problems that do not suit such description are those that involve objects that have length, area or volume, possess non-discrete coordinates, or the range of possible coordinates is not bound by some known value $a$. Although this excludes a large number of problems, there still remains some that fit in the given conditions. For example, in \cite{benitez2010parallel} proteins are represented as a chain of beads in $\mathbb{Z}^3$, so each bead is located at a limited distance from the previous one, which limits the coordinates to some interval, therefore allowing us to define a space $\S^3$ for the problem.

%\rev{Should we give the main insight of the algorithm before describing it?}

If the problem at hand can be made to fit into the aforementioned conditions, then the brute-force counting algorithm can be replaced by one with lower time complexity. First, the space $\S^3$ must be computationally represented in a way that each coordinate in $\S^3$ maps uniquely to a number stored in memory. This can be accomplished with a three-dimensional array of some numeric data type, which can be indexed using the coordinates of $\S^3$ themselves; however, to avoid indexing with negative numbers the space $\S^3$ must be translated to have its origin in $(a, a, a)$. For simplicity, in what follows we consider that three-dimensional arrays can be indexed with negative numbers.

\input{algorithm/1-collisioncount.tex}

Having defined the array that represents $\S^3$ in memory, we may then perform the counting algorithm.
% Assume, for now, that the array is initialized with zeros; the pattern of initialization depends on the description of the main body of the algorithm, so it is discussed later.
Algorithm \ref{algorithm:collision-count} shows the main procedure for counting, with linear time complexity, the number of collisions among a vector of beads, each of which has three integer coordinates. In this algorithm, the number associated with a point of $\S^3$ represents the number of beads in that spot so far  (assume, for now, that it is initialized with zeros). We begin by initializing the number of collisions with zero, and then iterate over the vector of beads. For each bead, we access the $\S^3$ space array using the bead's coordinates as index, retrieving from memory the number \textit{beadCnt} of beads in that place. We are on the process of adding a bead to a place that already contains \textit{beadCnt} beads, thus generating \textit{beadCnt} extra collisions that are added into \textit{collisions}. Finally, we increment the number associated with that place in the space, to effectively add one new bead there. For initializing the space array, we would also iterate over the vector of beads, initializing only elements at the coordinate of each bead.

% In order to function correctly, the \textit{space} array has to be initialized. The pattern of initialization can be derived by observing the set of elements that are read or written during the counting. As can be seen in Algorithm \ref{algorithm:collision-count}, we only access the \textit{space} array using the coordinates of each bead, so initialization can be done by iterating over the vector of beads a second time, as shown in Algorithm \ref{algorithm:memory-init}.

% \input{algorithm/2-meminit.tex}
\input{algorithm/3-contactcount.tex}

With another similar algorithm (see Algorithm \ref{algorithm:contact-count}), we can also calculate the total number of contacts among beads, that is, count how many pairs of beads are neighbors. As in the previous algorithm, the \textit{space} array holds how many beads are placed in each coordinate of $\S^3$ so far, and we begin by initializing a \textit{contacts} variable with zero. The first loop ``places'' beads in the \textit{space} array, such that its element $(x,y,z)$ has the number of beads with coordinates $(x,y,z)$. Afterwards, we iterate over the vector of beads again, and for each bead \textit{b} we fetch from \textit{space} the number of beads in each of the six spots that are neighbors of \textit{b}, and add them into \textit{contacts}. There is still a problem to deal with: if beads $b_1$ and $b_2$ are neighbors, their contact is counted twice (in iterations for $b_1$ and $b_2$); because of this, the function returns \textit{contacts/2}.

Determining the initialization pattern for the problem of contact counting follows a reasoning similar to the one used with collisions. The elements of \textit{space} that are accessed are all neighbors of each bead, including the bead's own position, so they must be initialized to zero.

The presented algorithms involve 2 steps: initializing \textit{space} and counting either contacts or collisions. Both steps consist of $N$ iterations, one for each bead, and in each iteration we perform a fixed number of $O(1)$ instructions: sum, subtraction and memory load/store. Hence, the algorithms have $O(N)$ time complexity, so they tend to perform better than the quadratic approach for large enough $N$. However, experiments show that these algorithms are faster even for small $N$ ($< 64$), which are elaborated in Section \ref{section:5-experiments}.

On the other hand, the algorithms make use of the three dimensional array \textit{space}, whose size depends on the cardinality of $\S^3$. As defined earlier, $\S = \{-a,\ -a+1,\ \dots,\ a-1,\ a\}$, so each axis of $\S^3$ has $2a+1$ elements, giving a total of $(2a+1)^3$ elements. Consequently, the algorithms presented have $O(a^3)$ complexity of memory consumption. The $a$ might be known on compilation time, in problems where the beads are confined in a box of known edge length, regardless of the number $N$ of beads. Another possibility is that $a$ is a function of $N$; for example, take the case of proteins that are modelled as a chain of beads that begins in $(0,0,0)$, a chain of size $N$ would require each axis to span from $-N$ to $N$, meaning $a$ would be a function $a(N) = N$, and the memory usage complexity can be rewritten as $O(N^3)$.

Although there are cases in which the memory complexity is $O(N^3)$, this concerns only virtual memory. When counting collisions, for example, virtual pages are mapped physically only if they contain beads. In the worst case each bead will be in a different page and $N$ pages will be mapped, making it $O(N)$ in physical memory usage. An important consequence of this is that swap memory tends to be depleted before physical memory, so by the time swap depletes the counting program still will not have begun to access disk for swapping, which would greatly impact performance.

% Restore the old \S command
\let\S\oldS

%% file: algorithm/1-collisioncount.tex
\begin{algorithm}[htb]
    \caption{Counting number of collisions among a vector of beads.}
    \label{algorithm:collision-count}
    
    % \footnotesize
    
    \SetStartEndCondition{ (}{)}{)} % Open and close conditions with ()
    \SetAlgoBlockMarkers{}{\}}      % Close blocks with }, but don't open with {, which are set up on the \SetKw
    \AlgoDisplayBlockMarkers        % Display the { }
    \SetKwFor{For}{for}{\{}{}       % Define a for block
    %\SetKwIF{If}{ElseIf}{Else}{if}{\{}{else if}{else\{}{}   % Define an if...else if...else block
    \SetKwFor{While}{while}{\{}{}   % Define a while block
    \SetArgSty{textnormal}          % Remove italic font from conditions
    \SetFuncArgSty{textnormal}      % Remove italic font from function arguments
    \SetFuncSty{textit}             % Set function names as italic
    %\SetInd{0.4em}{0.3em}           % Set indentation
    %\Indm\Indm                      % Remove indentation of line numbers

    \SetKwProg{countFunc}{int}{\{}{}
    \SetKwFunction{countCol}{countColls}
    \SetKw{In}{in}
    \SetKwIF{If}{ElseIf}{Else}{if}{}{else if}{else}{}

    \countFunc{\countCol{\textbf{point3D} beads[], \textbf{int} space[][][]}}{
        \textbf{int} collisions = 0\;
        \For{b \In beads}{
            \textbf{int} beadCnt = space[b.x][b.y][b.z]\;
            collisions += beadCnt\;
            space[b.x][b.y][b.z] += 1\;
        }
        \textbf{return} collisions\;
    }
\end{algorithm}

%% file: algorithm/3-contactcount.tex
\begin{algorithm}[htb]
    \caption{Counting number of contacts among a vector of beads.}
    \label{algorithm:contact-count}
    
    %\footnotesize
    
    \SetStartEndCondition{ (}{)}{)} % Open and close conditions with ()
    \SetAlgoBlockMarkers{}{\}}      % Close blocks with }, but don't open with {, which are set up on the \SetKw
    \AlgoDisplayBlockMarkers        % Display the { }
    \SetKwFor{For}{for}{\{}{}       % Define a for block
    \SetKwIF{If}{ElseIf}{Else}{if}{\{}{else if}{else\{}{}   % Define an if...else if...else block
    \SetKwFor{While}{while}{\{}{}   % Define a while block
    \SetArgSty{textnormal}          % Remove italic font from conditions
    \SetFuncArgSty{textnormal}      % Remove italic font from function arguments
    \SetFuncSty{textit}             % Set function names as italic
    %\SetInd{0.4em}{0.3em}           % Set indentation
    %\Indm\Indm                      % Remove indentation of line numbers

    \SetKwProg{countFunc}{int}{\{}{}
    \SetKwFunction{countConts}{countContacts}
    \SetKw{In}{in}
    \SetKwFor{ForTwo}{for}{}{}

    \countFunc{\countConts{\textbf{point3D} beads[], \textbf{int} space[][][]}}{
        \textbf{int} contacts = 0\;
        \lForTwo{b \In beads}{\\\hspace{0.5em}space[b.x][b.y][b.z] += 1}
        \For{b \In beads\label{algorithm:l1-contact-count}}{
            contacts += space[b.x+1][b.y][b.z]\;
            contacts += space[b.x-1][b.y][b.z]\;
            contacts += space[b.x][b.y+1][b.z]\;
            contacts += space[b.x][b.y-1][b.z]\;
            contacts += space[b.x][b.y][b.z+1]\;
            contacts += space[b.x][b.y][b.z-1]\;
        }
        \textbf{return} contacts / 2\;
    }
\end{algorithm}

%% file: text/4-parallelalgorithm.tex
\section{An Efficient Parallel Algorithm}
\label{section:4-parallelalorighms}

% The $O(N)$ approach tends to incur high memory consumption, and is not viable for large problem sizes, in which case it might be reasonable to use the $O(N^2)$ approach instead. Such quadratic approach may be subject to parallelization, which can yield great performance gains. In the following, we discuss two efficient parallelizations for the quadratic algorithm; but first, we need to further analyze the sequential algorithm.
The $O(N)$ approach tends to incur high memory consumption, which poses an obstacle to handle large problem sizes, in which case it is reasonable to use parallel computing to distribute memory usage or computation among processing nodes. Applying this technique is possible for both the $O(N)$ and the $O(N^2)$ approaches, but due to the seemingly higher difficulty in using it for the $O(N)$ one, we analyze and propose an efficient parallelization of the $O(N^2)$ algorithm for counting any kind of symmetric pairwise interactions (SPI) among objects, not limited to collisions. The proposed algorithm is aimed mainly at GPUs, since it makes better use of its architectural characteristics, and experiments were performed using them. However, results are not limited to GPUs as there might be parallel architectures with similar characteristics, and could be better exploited with the algorithm presented in the following.

In Algorithm \ref{algorithm:sequentialcode}, we present the standard sequential code for calculating SPI. For each object, we accumulate its interaction with all subsequent objects. Analyzing the nested loops in search for parallelization, we notice that they are not completely data-parallel due to the \textit{interactions} variable, which is read and written in every iteration. Such variable is a reduction variable, which implies that parallelizing the iterations is still viable, provided that there is a reduction phase that agglomerates the intermediate results calculated by each parallel execution unit (denoted as \textit{threads} from here on).

A second aspect of the standard sequential algorithm is that the outer loop is not balanced in terms of work executed per iteration. The first outer iteration executes $N-1$ inner iterations, whereas the last outer iteration executes none. When we try to parallelize the problem, this could cause threads to be assigned different amounts of work, resulting in idle threads waiting for others to finish their larger burden. One way to promote balancing is to assign outer iterations to GPU threads in a round-robin fashion such that each thread performs at least two outer iterations. However, by agglomerating outer iterations this approach reduces the number of threads we can launch, and it also arguably reduces memory locality, which are undesirable properties for GPU algorithms.
% This can be avoided by assigning iterations to threads in a round-robin fashion, but this might impact memory locality because the pattern by which threads read objects from memory will not be contiguous anymore, they will skip over memory words according to a certain stride.

\vspace{-0.3cm}
\input{algorithm/4-sequentialalgorithm.tex}
\vspace{-0.8cm}
\input{algorithm/5-proposedalgorithm.tex}
\vspace{-0.3cm}

For balancing the loop iterations in a way optimized for GPU, we offer an alternative solution whose main portion of code is shown in Algorithm \ref{algorithm:proposedcode}. We now explain the idea behind it, prove that it works when N is odd and finally elaborate on how to make it work with even N too.

In the proposed algorithm, the outer loop can be seen as follows. Each bead $i$ (we will use the term \textit{bead} for simplicity, but they can be any kind of object) evaluates the interaction of itself with 
%some 
subsequent beads
%. This happens 
in a circular fashion.
%, such that if the current bead evaluates the last bead, on the next iteration it continues from the beginning of the vector.
%Considering that we have $N$ beads numbered from $0$ to $N-1$, the circularity can be mathematically modelled by working with integers \textit{modulo N} \cite{knuth1989concrete}. In this universe ($\mathbb{Z}_N$), numbers whose division by $N$ yield the same remainder are considered equivalents, that is, $1 \equiv_N N+1$, $2 \equiv_N N+2$ and so on; note that if we count beads as $0, 1, 2, ...$, we eventually reach $N$ which is equivalent to $0$ modulo $N$; this effectively models the circularity without having to reset the counting back to $0$.
This circularity can be mathematically modelled by working in the universe of integers \textit{modulo N} \cite{knuth1989concrete}, which has interesting properties that we will use later:
\begin{align}
    a \equiv_N b \implies &\forall k \in \mathbb{Z}\ \ \ a + k \equiv_N b + k \label{equation:1} \\
    %a \equiv_N b \implies &\forall k \in \mathbb{Z}\ \ \ k.a \equiv_N k.b \\
    a \equiv_N 0 \iff &\exists c \in \mathbb{Z}\ \ \ a = c.N \label{equation:2}
\end{align}

\begin{figure}[htb]
    \centering
    \includegraphics[width=0.7\linewidth]{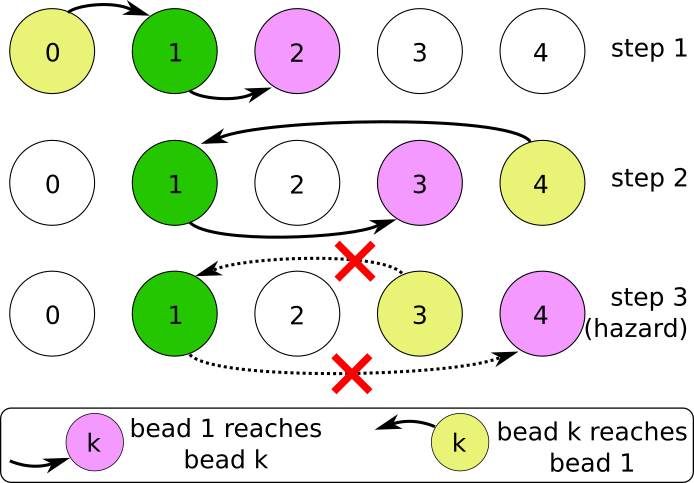}
    \caption{Algorithm execution from the perspective of bead 1. Bead 1 reaches beads 2 and 3, and it is reached by beads 0 and 4 as $s$ increases.}
    \label{figure:beadscrossing}
%\vspace{-0.4cm}
\end{figure}

For each bead $i$ we can now define two functions: $reach(s)$, that returns the index of the bead being evaluated in step $s$; and $reached(s)$ that returns the index of the bead evaluating bead $i$ in step $s$. The algorithm begins in step $s = 1$ and goes forward until some stopping condition that we discuss now. For any bead $i$, we have:
%\vspace{-0.25cm}
\begin{center}
    \begin{tabular}{ c c c }
         \hspace{0.1cm}$s = 1$\hspace{0.1cm} &  \hspace{0.1cm}$reach(1) \equiv_N i + 1$\hspace{0.1cm} & \hspace{0.1cm}$reached(1) \equiv_N i - 1$\hspace{0.1cm} \\
         $s = 2$ & $reach(2) \equiv_N i + 2$ & $reached(2) \equiv_N i - 2$ \\
         & ... & \\
         $s$ & $reach(s) \equiv_N i + s$ & $reached(s) \equiv_N i - s$ \\
    \end{tabular}
%\vspace{-0.2cm}
\end{center}
where bead $i$ evaluates beads $j$ with $j$ increasing, and bead $i$ is evaluated by bead $k$ with $k$ decreasing as $s$ advances (see Figure \ref{figure:beadscrossing}).

%\vspace{-0.4cm}

An undesired situation here is that bead $i$ evaluates bead $j$ when bead $j$ has already evaluated bead $i$, situation in which we would be evaluating the same interaction twice. This is illustrated in Figure \ref{figure:beadscrossing}, where this situation happens in step 3, where bead $3$ reaches bead $1$, but bead $1$ had already reached bead $3$ in step 2. For any odd number $N$ of beads, this hazard happens on the the step where bead $i$ reaches bead $j+1$ on the same step that bead $j$ reaches bead $i$. Mathematically, the violation happens when $reach(s) - 1 \equiv_N reached(s)$, which for an arbitrary bead $i$ gives
\vspace{-0.22cm}
\begin{equation}
\begin{split}
    i + s - 1 &\equiv_N i - s \\
    2s - 1 &\equiv_N 0 \hspace{2.44cm} \text{\textbf{using (\ref{equation:1})}} \\
    2s - 1 &= c.N \hspace{1cm} c \in \mathbb{Z} \hspace{0.5cm} \text{\textbf{using (\ref{equation:2})}} \\
    s &= \frac{c.N + 1}{2}\ \therefore\ s = \frac{N + 1}{2}
\end{split}
\end{equation}

Note that modular arithmetic gives a set of answers instead of a unique one. It shows that the step $s$ where the violation occurs is $(c.N + 1) / 2$ for some integer $c$. If $c$ is $0$, then $s = 1/2$ which is invalid because, as stated earlier, the first step is $s = 1$; any $c < 0$ gives $s < 0$ which also does not make sense; $c = 1$ gives $(N + 1) / 2$, which is in fact the first step where the violation occurs ($N$ is odd, so the division results in an integer). Taking $c = 2, 3,$ and so on yields higher steps where the violation would occur, but they do not matter for us because we will stop the algorithm before the first violation. Finally, since the first violation occurs at $s = (N+1)/2$, we need to stop at the previous iteration, that is, at $s = (N-1)/2$.

We still need to prove the equivalence between the proposed and straightforward algorithms. First, the number of different interactions among beads amount to $N.(N-1)/2$, which is precisely the amount of interactions evaluated by the straightforward approach. We now prove that the proposed algorithm performs this same amount of work, and that all interactions evaluated are mutually different, hence proving the equivalence of both approaches.

% \vspace{1em}
\begin{elfproposition} \label{prop:same-work}
\it The proposed algorithm evaluates $N.(N-1)/2$ interactions among beads, for odd $N$.
\end{elfproposition}
\noindent \textit{Proof.}\ \ As stated before, each of the $N$ beads evaluates its interaction with $(N-1)/2$ subsequent beads, which amounts to a total of $N.(N-1)/2$ evaluations. \qed

% \begin{proposition} \label{prop:same-work}
% The proposed algorithm evaluates $N.(N-1)/2$ interactions among beads, for odd $N$.
% \end{proposition}
% \noindent \textit{Proof.} As stated before, each of the $N$ beads evaluates its interaction with $(N-1)/2$ subsequent beads, which amounts to a total of $N.(N-1)/2$ evaluations.

% \vspace{1em}
\begin{elfproposition} \label{prop:diff-interacts}
\it All $N.(N-1)/2$ interactions evaluated by the proposed algorithm are different from each other, for odd $N$.
\end{elfproposition}
\noindent \textit{Proof.}\ \ Take two arbitrarily different beads $i$ and $j$, with $i < j$. Each bead evaluates the interaction between itself and subsequent beads, so for beads $i$ and $j$ one side of the interactions they evaluate is inherently different, since $i \ne j$. Consequently, both beads would only evaluate the same interaction if it was the interaction among beads $i$ and $j$ themselves. Let us see when this happens.

For bead $i$ to evaluate interaction of $i$ with $j$, it would be necessary that
\begin{align*}
    reach(s) &\equiv_N j \hspace{1cm} \text{from bead i's perspective} \\
    i + s &\equiv_N j \\
    i + s - j &\equiv_N 0 \tag{\stepcounter{equation}\theequation}\\
    i + s - j &= c.N \hspace{1cm} c \in \mathbb{Z} \\
    s &= c.N + (j - i)\ \therefore\ s = j - i
\end{align*}
For $c < 0$, $s$ would be a negative value because $j - i$ (the distance between beads $i$ and $j$) cannot be higher than $N-1$, and this is a contradiction because negative $s$ never happens. If $c > 0$, then $s = c.N + (j - i) > N$, but $s$ only gets values $1, ..., (N-1)/2$. Finally, taking $c = 0$ makes $s = j - i$, which happens as long as $j - i$ also resides in interval $1, ..., (N-1)/2$, that is, the distance between $j$ and $i$ is lower than or equal to $(N-1)/2$. On the other side, bead $j$ will evaluate its interaction with $i$ when its $reach(s) \equiv_N i$, and developing this expression in a similar way as before we obtain $s = c.N - (j - i)$, but the only valid value for $c$ is $1$. This gives $s = N - (j - i)$, and $s \in \{1, ..., (N-1)/2\}$ only if $j - i \ge (N+1)/2$.

% \begin{equation}
% \begin{split}
%     reach(s) &\equiv_N i \hspace{1cm} \text{from thread j's perspective} \\
%     j + s &\equiv_N i \\
%     j + s - i &\equiv_N 0 \\
%     j + s - i &= c.N \hspace{1cm} c \in \mathbb{Z} \\
%     s &= c.N + (i - j) \\
%     s &= N - (j - i) 
% \end{split}
% \end{equation}
% For $c \le 0$, we would again try to make $s$ be negative, which will not ever happen. For $c > 1$, we would try to compare $s$ with values higher than $N$, also invalid. Finally, for $c = 1$, we have $s = N - (j - i)$, which only happens if $j - i$, the distance between $j$ and $i$, is greater than or equal to $(N+1)/2$.

Therefore, let the distance between $i$ and $j$ be called $d = j - i$, then $i$ will evaluate $j$ only if $d \le (N-1)/2$, and $j$ will evaluate $i$ only if $d \ge (N+1)/2$. This means that $i$ and $j$ do not mutually evaluate each other, so the interaction among beads $i$ and $j$ is evaluated only once. This proves that all interaction evaluations are mutually different, completing the proof of equivalence between this approach and the straightforward one, for odd $N$. \qed

When $N$ is even, a slight modification is needed\footnote{\label{footnote1}The full formulation, as well the source code for all experimented programs mentioned in this article, is available in \href{https://mjsaldanha.com/articles/1-hpc-sspi/}{mjsaldanha.com/articles/1-hpc-sspi/}.}. The stopping condition of the outer loop is derived in a similar way as done before, but in this case the problem is not that the $reach()$ and $reached()$ arrows cross themselves; instead, they reach the same value. That is, for a given bead $i$ we have $reach(s) \equiv_N reached(s)$, which will result, following the same mathematical steps as before, in $s = N/2$. This is the step in which bead $i$ is evaluating some bead $j$ while bead $j$ is also evaluating bead $i$. To prevent this situation, we allow execution of $N/2 - 1$ steps normally, and the first half of the beads are made to execute one more. This works because in step $N/2$ the beads being reciprocally evaluated are on different halves of the vector of beads, since they are within a distance of $N/2$ from each other. The consequence of this modification is that the algorithm is not completely balanced any longer; some outer iterations execute one extra inner iteration.

This concludes the formulation of the alternative, balanced algorithm. This proposed approach has some nice theoretical properties, based on the concepts of \textit{depth} and \textit{work} \cite{blelloch1996programming}. Depth is the largest amount of work done sequentially by a single thread, while work is the total amount of work done by all threads launched. In the straightforward algorithm, we have a work-complexity of $N.(N-1)/2$ because that is the number of interactions that need to be evaluated, and a depth-complexity of $N-1$ because the thread that performs the first outer loop evaluates that many interactions. Note that we are ignoring the work and depth of the reduction phase because it is performed in the exact same way in both the straightforward and the proposed approaches. In our proposed algorithm, the work-complexity is maintained (seen in Proposition \ref{prop:same-work}), while the depth-complexity becomes $(N-1)/2$ for odd $N$, and $N/2$ for even $N$. Therefore, we reduced the depth while preserving the amount of work, which indicates that if we had infinite physical processing units, the proposed approach could be faster.

In practice, both approaches can be parallelized by assigning each outer iteration to one thread, followed by a reduction phase where the threads cooperate to accumulate the intermediate results obtained. As was already mentioned, it is possible to parallelize the straightforward approach in a balanced way by distributing the outer loops over a smaller number of threads in a round-robin fashion. However, although this might perform well in distributed or multicore systems, it increases the algorithm's depth, reduces the number of threads that can be launched and degrades memory locality, which aren't good properties for GPUs. Besides that, it also makes it considerably more difficult to manage usage of shared memory, which is a fast memory shared only by a block of threads. For these reasons, we have implemented in the CUDA programming model only the straightforward parallelization and the proposed one, and we show in Section \ref{section:5-experiments} that the proposed approach is slightly better.

% Maybe place some emphasis on the complexity of \textit{interacts}. If it is highly complex, our algorithm is certain to give nice improvements if we have infinite processors.

%% file: algorithm/4-sequentialalgorithm.tex
\begin{algorithm}[htb]
    \caption[caption]{Standard algorithm for calculating SPI.}
    \label{algorithm:sequentialcode}

    % \footnotesize

    \SetStartEndCondition{ (}{)}{)} % Open and close conditions with ()
    \SetKwFor{For}{for}{}{}         % Define a for block
    \SetKwIF{If}{ElseIf}{Else}{if}{}{else if}{else}{}   % Define an if...else if...else block
    \SetKwFor{While}{while}{}{}     % Define a while block
    \SetArgSty{textnormal}          % Remove italic font from conditions
    \SetFuncArgSty{textnormal}      % Remove italic font from function arguments
    \SetFuncSty{textit}             % Set function names as italic
    %\SetInd{0.2em}{0.3em}           % Set indentation
    %\Indm\Indm                      % Remove indentation of line numbers

    \SetKwFunction{int}{interact}
    \SetKw{In}{in}
    \SetKw{To}{to}

    \For{i = 0 \To N-1}{
        \For{j = i+1 \To N-1}{
            interactions += \int{obj[i], obj[j]}\;
        }
    }
\end{algorithm}

%% file: algorithm/5-proposedalgorithm.tex
\begin{algorithm}[htb]
    \caption[caption]{Proposed algorithm for calculating SPI.}
    \label{algorithm:proposedcode}

    % \footnotesize

    \SetStartEndCondition{ (}{)}{)} % Open and close conditions with ()
    \SetKwFor{For}{for}{}{}         % Define a for block
    \SetKwIF{If}{ElseIf}{Else}{if}{}{else if}{else}{}   % Define an if...else if...else block
    \SetKwFor{While}{while}{}{}     % Define a while block
    \SetArgSty{textnormal}          % Remove italic font from conditions
    \SetFuncArgSty{textnormal}      % Remove italic font from function arguments
    \SetFuncSty{textit}             % Set function names as italic
    %\SetInd{0.2em}{0.3em}           % Set indentation
    %\Indm\Indm                      % Remove indentation of line numbers

    \SetKwFunction{int}{interact}
    \SetKw{In}{in}
    \SetKw{To}{to}
    \SetKw{Int}{int}
    \SetKw{Even}{is even}

    \For{i = 0 \To N-1}{
        \For{j = 1 \To (N-1)/2}{
            interactions += \int{obj[i], obj[(i+j)\%N]}\;
        }
        % j = N/2 + 1\tcc*{an extra iteration}
        % \If{N \Even \&\& i $<$ N/2}{
        %     \lIf{\int{beads[i], beads[ (i+j)\%N ]}}{interactions++}
        % }
    }
\end{algorithm}

%% file: text/5-experiments.tex
\section{Experiments and Results}
\label{section:5-experiments}

In order to evaluate the performance of the $O(N)$ sequential counting algorithm of Section \ref{section:3-linearapproach}, we implemented both approaches, linear and quadratic, for counting collisions\textsuperscript{\ref{footnote1}}.
% Since the $O(n)$ approach is sensitive to how it is used, 
We designed the implementations so that they had similar characteristics to the protein structure prediction program (from \cite{benitez2010parallel}) analyzed and implemented in the broader context of this research. Hence, in each execution we perform the counting procedure upon multiple bead vectors, the space array is allocated only once, and each bead vector is generated by placing the first bead at $(0,0,0)$ and positioning the next bead in the neighborhood of the previous bead (similar to a protein), choosing any of the 6 directions randomly.

\begin{figure*}[t]
    \centering
    \includegraphics[width=0.4\linewidth]{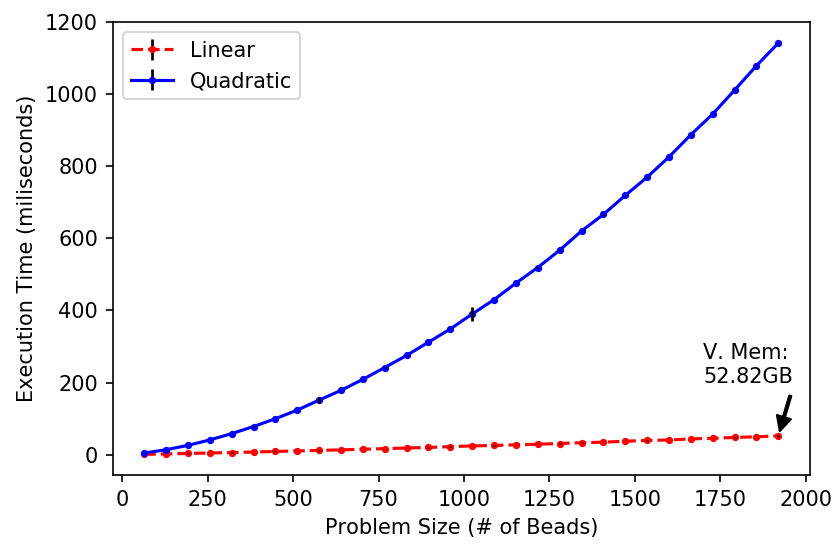}
    \includegraphics[width=0.4\linewidth]{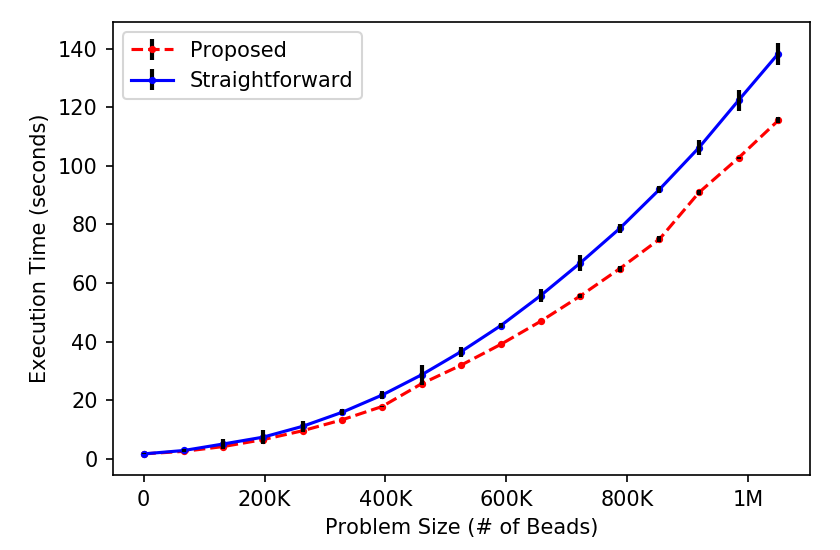}
    \vspace{-0.4cm}
    \caption{To the left, execution time for linear and quadratic approaches for counting collisions. To the right, for counting collisions in GPU. All vertical black error bars have length of four standard deviations of 100 samples taken.}
    \label{fig:experiments-joint}
    %\vspace{-0.4cm}
\end{figure*}

Figure \ref{fig:experiments-joint} (left side) shows the experimental results. Each program was executed with varying problem sizes and each execution comprised counting the number of collisions for 1000 different bead vectors. For each problem size we collected the wall clock time for each of 100 executions and took their mean. The total vertical length of the black error bars equals four standard deviations of the samples. These experiments were run using an Intel i7-4790 3.6GHz and 32GB of primary memory. For a problem size of 1920, which was the largest problem size that could be run in the system, the speedup was 21.7 and the linear approach required 52.82GB of virtual memory, as pointed in Figure \ref{fig:experiments-joint} (left).

% \begin{figure}[htb]
%     \centering
%     \includegraphics[height=6cm]{img/fig_psp_whole.png}
%     \caption{Execution time for the protein structure prediction program using quadratic and linear approaches for counting collisions and contacts. Each vertical black error bar has length of four standard deviations.}
%     \label{figure:exp-psp-whole}
% \end{figure}

With the results shown in Figure \ref{fig:experiments-joint} (left), the gain in execution time provided by the linear approach is clear. However, this comes at the cost of high consumption of virtual memory, which greatly limits the largest problem size that can be supported by one's system. 
% Matheus, rever esta frase abaixo. Está confuso: expectativa do nr de beads & programa implementado & necessidade de se falar de PCs.
In the case of the protein structure prediction (PSP) program we implemented\footnotemark[2] during the research project that revolve this paper, the number of beads among which collisions are calculated rarely exceeds 1000 (proteins rarely have that number of amino acids), and for this problem size the virtual memory usage is 7.5 GB, a feasible amount.
% Matheus, rever esta frase abaixo. Qual a relação das 50 execuções com o resultado apresentado?
Experiments with the PSP program showed speedups of 11.8 and 72.4 for proteins with 128 (a common size) and 768 aminoacids, respectively, so by using the proposed algorithm the program was accelerated significantly. Possible reasons for the higher speedup are related to factors that are discussed below.
% Experiments with the PSP program\footnotemark[2], in similar conditions as the counting algorithms, is shown in Figure \ref{figure:exp-psp-whole} and show how faster the prediction became when using linear approaches for counting collisions and contacts. Here we executed the programs 50 times for each problem size, and the error bars again show four standard deviations.

\begin{figure*}[t]
    \centering
    \includegraphics[width=0.4\linewidth]{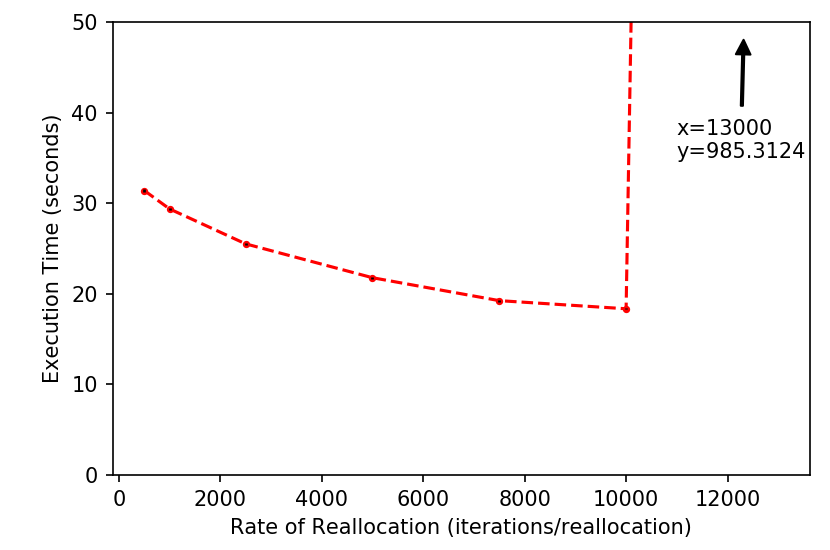}
    \includegraphics[width=0.4\linewidth]{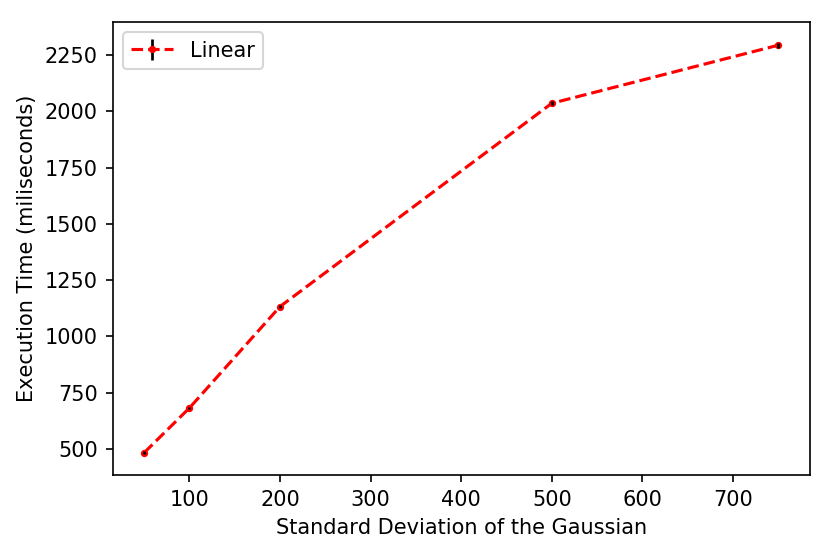}
    \vspace{-0.4cm}
    \caption{Linear approach when: (to the left) varying the rate by which memory is reallocated, for a problem size of $13\,000$ vectors of beads; and (to the right) generating bead coordinates with a normal distribution with mean $0$ and varying standard deviation, for a problem size of $1000$ bead vectors.}
    \label{fig:experiments-considerations}
    %\vspace{-0.6cm}
\end{figure*}

Some considerations must be made regarding the generation of beads in resemblance to the PSP algorithms. By allocating the space array only once in each execution, we reduce the cost of requesting memory from the operating system, which is present only in the linear approach; on the other hand, this means we do not reclaim virtual memory after using it for a single vector of beads, causing physical pages that were mapped to remain mapped until the end of the program execution.
For bigger problem sizes, memory should be reallocated every $K$ iterations so as to free unused allocated virtual pages and prevent the program to swap memory. In Figure \ref{fig:experiments-considerations} (left), we show what happens to the execution time when varying such number $K$ of iterations. Notice that performance improves as we reallocate memory less often (due to lower memory management overhead), reaching an optimal point at a rate of one reallocation every $10\,000$ iterations. After that, the program begins to swap memory and performance degrades significantly.

% Matheus, precisa explicar melhor a questão abaixo. 
A second consideration is that, by generating bead vectors as ``random proteins'', we are statistically confining beads to a smaller region around the origin of the space, hence improving cache usage and reducing the number of memory pages required.
For better illustrating this, we show in Figure \ref{fig:experiments-considerations} (right) what happens to execution time if we generate the beads' coordinates using a normal distribution with mean $0$ and varying standard deviation. As expected, performance degrades at higher deviation due to lower rate of cache hits.
%and then increases due to truncation of beads to fit inside the space $\mathcal{S}$ (see Section \ref{section:3-linearapproach}).
% (performance then increases because too many beads fall outside of the space $\mathcal{S}$, defined in Section \ref{section:3-linearapproach}, and are ``truncated'' to be inside $\mathcal{S}$).
% In the PSP program, the optimization algorithm favors compact proteins, which further decreases the memory pages required and possibly allows for better cache usage, which explains the difference in the aforementioned speedups (72.4 compared with 21.7).

It follows from these two considerations that the frequency of reallocation of the space array, as well as the regularity of bead positions, must be taken into consideration when using the linear approach. There is a lot of room for analyzing such factors and how they apply to real applications; as that is not the objective of this paper, it is left as future work.

%\vspace{-0.5cm}

The parallel algorithms for SPI presented in Section \ref{section:4-parallelalorighms} were implemented\footnotemark[2] in the CUDA programming model and evaluated using NVIDIA GPUs.
Experiments used a NVIDIA Tesla P100 with 16GB of memory and 3584 CUDA cores spread over 56 multiprocessors.
Both the straightforward and the proposed approaches were optimized to achieve high occupancy and memory bandwidth, and all optimizations were applied to both approaches alike. Then, we executed each program 100 times for each problem size and took the mean of their wall clock execution times, which are shown in Figure \ref{fig:experiments-joint} (right side); four standard deviations are represented by the black vertical error bars. Each program execution comprised calculating collisions for 100 randomly generated vectors of floating point spheres with diameter of 1. Even though comparisons with sequential approaches could have been made, it is quite clear that parallel algorithms yield lower execution times and the objective here is mainly to show the benefits of the proposed parallelization compared with the straightforward one.
%and the CUDA kernels for performing the counting were launched one at a time, that is, they were not launched to execute in parallel.

Figure \ref{fig:experiments-joint} (right) shows that the proposed approach seems to perform well on GPU. In fact, for all problem sizes larger than $525\,000$, the proposed approach is more than $12\%$ faster than the straightforward one ($p < 0.01$ using a t-test assuming unknown and different variances). 
The speedup is mainly due to two factors. First, in the straightforward approach each CUDA block is responsible for calculating interactions of a group of 1024 spheres with \textit{all} subsequent spheres; because of this, some blocks perform more work than others, and in the last moments not all of the 56 multiprocessors are used, and the program is waiting for a few lengthy blocks to end their part of the work. 
Second, besides this block-level unbalance, threads within a warp (group of 32 threads that execute simultaneously) are also unbalanced, so when the warp is nearly concluding its work the 32\textsuperscript{nd} thread ends its job earlier and becomes idle waiting for the other threads to finish. 
In our proposed balanced approach, threads or multiprocessors that would otherwise be idle are put to perform useful work and contribute to finish more quickly the interaction counting. Finally, if the problem in hand allowed the CUDA kernels to be launched in parallel, the GPU could hide a lot of the block-level unbalance; however, the warp-level one would remain negatively impacting the program speed.
%
%Matheus, precisa comparar os resultados das versões paralelas com as versões sequenciais. 
% Precisa comparar também o impacto desse cálculo nos algoritmos que necessitam dele. Por exemplo, no caso desse cálculo dos beads ser feito a cada iteração desses algoritmos, o impacto final dessa diferença de 12% é grande.
%Enfim, falta mostrar melhor o impacto dos benefícios alcançados pela pesquisa feita.

%% file: text/6-conclusions.tex
\section{Conclusion}
\label{section:6-conclusion}

Interaction counting is a usual problem, and when it needs to be performed, the quadratic pairwise-comparisons approach immediately comes to mind. For a long time this has been a significant bottleneck\cite{lin1998collisionsurvey} on important areas such as computer graphics and scientific simulations; in the former, collision counting must be performed enough times per second to allow image frames to be delivered in a visually fluid way, and more frames mean more fluidity, so every millisecond matters; in the latter, simulations of galaxies or proteins may need a large number of iterations if a high level of reality is desired (possibly taking weeks to execute), so if the interaction counting performed every iteration is accelerated, either less simulation time would be required or more iterations could be performed in order to achieve better results. In either case, performing interaction counting in less time yields great benefits, which is why a lot of research has been carried out on the subject. However, research walked toward algorithms that focus on reducing the number of objects among which interaction counting must be performed using the usual $O(N^2)$ approach, often in parallel.

In this paper are proposed two algorithms that aim at improving the pairwise-comparisons approach itself: a sequential approach with $O(N)$ complexity that works well for punctual objects in a limited discrete space, and a parallel approach that runs more efficiently on GPUs than the pairwise-comparisons algorithm's straightforward parallelization. These approaches can, of course, be used together with the algorithms that focus in pruning sets of interacting objects, in order to accelerate the phase where brute-force interaction counting must be performed. By using the $O(N)$ approach, interaction counting can be made significantly faster, at the cost of high memory consumption; our experience with accelerating a protein structure prediction algorithm, as already mentioned, shows a speedup of 72.4 using the same hardware. The proposed parallel algorithm may be used on large problems, with objects of any shape, for counting any kind of interactions, and experiments using GPU show that it can yield a 1.12 speedup.

% I found an explanation for the usage of "every which" here: http://www.englishtown.com/EtownResources/Grammar/27.html
A possible direction for future research is to evaluate possible benefits of parallelizing the proposed $O(N)$ algorithm in order to share the memory consumption among nodes, which would allow the algorithm to be used for bigger problem sizes. Also, the $O(N)$ algorithm is sensitive to cache effects and to how frequently memory is allocated and deallocated, so these factors should be further investigated, especially when applied to real problems.
%; considering the goals of minimizing swap disk access and time spent with memory management system calls, 
%there probably exists an optimal $K$ such that memory should be reallocated every $K$ iterations.
%there probably exists an optimal number of iterations every which memory should be reallocated.
Besides that, the proposed parallel algorithm should be implemented for different parallel platforms that could have their architectural resources better used by the algorithm, since it has nice properties; examples of platforms are Intel Xeon Phi processors and FPGAs.

%% file: text/9-acknowledgements.tex
\section*{Acknowledgement}

% Tirado de: http://agencia.fapesp.br/fapesp-estabelece-normas-para-o-reconhecimento-de-projetos-apoiados-/17484/

We thank São Paulo Research Foundation (FAPESP) for funding this research project (grant 2017/25410-8, associated to 2013/07375-0), and the Center for Mathematical Sciences Applied to Industry (CeMEAI) for providing access to powerful computational resources.